\documentclass[prl,groupaddress,showpacs,twocolumn]{revtex4}
\usepackage{graphicx}

\begin{document}

\title{Yet another hysteresis model}

\author{Sergey Langvagen}
\email{lang@ezan.ac.ru}
\affiliation{Chernogolovka, Moscow region, Russia}

\date{\today}

\begin{abstract}
A hysteresis model based on the assumption of fixed order magnetization reversals is proposed.
The model  uses one-dimensional diagram for representing states of a system despite of
two-dimensional Preisach diagram. 
The distinctive feature of the model is that it is applicable to any  system compliant with the return-point memory
and includes Preisach model as a special case.

\end{abstract}


\maketitle

\section{Introduction}

The problem of hysteresis modeling is a subject of persistent  interest.
One of the most popular, Preisach model,  was proposed more than 65 years ago \cite{Preisach}. 
It was initially based on some hypothesis concerning physical mechanisms of magnetization,  
but now it is considered mainly as a mathematical tool for describing
various hysteresis phenomena. 
Preisach model has many attractive features such as simplicity, flexibility and ability to represent most important 
properties of real hysteresis systems \cite{Mayergoyz, Bertotti, DellaTorre}.  
There are known numerous applications of Preisach model in different areas of physics.
The postulates of Preisach model lead  to  the return-point memory   \cite{Sethna&all1993},  the property of  real systems with hysteresis
that is  considered as one of the most sufficient for hysteresis modeling.

In this article we intoduce a simple hysteresis model, which we call {\em state vector} model.
It uses very different postulates but is similar to Preisach model in that
it is compliant with the return-point memory and can be easily analysed. 

\section{Description of the model}

Let us assume that states of a system can be represented by the ``state vector'' $x$ 
with components $x_\alpha$ equal to $+1$ or $-1$, where  $\alpha = 1 \ldots N$, and  $N$ is some large integer. 

Suppose that the magnetization $M$ (and other physical values related to the system) depend on $x$:
\begin{equation} \label{M}
M=M(x_1,\ldots , x_N).
\end{equation}

The behavior of the state vector $x$ is postulated as follows.

\begin{enumerate}
\item  When the magnetic field $H$ increases,  negative
components $x_\alpha$ change the sign from $-1$ to $+1$  one by one  in  order of increasing $\alpha$.
\item  When the magnetic field $H$ decreases, positive components $x_\alpha$ change the sign from
$+1$ to $-1$  one by one  in order of increasing $\alpha$.
\item Each state $x=(x_1,x_2,\ldots,x_N)$ of the system has a value of the
external magnetic field
\begin{equation}
H=H(x_1,\ldots,x_N)  \label{H}
\end{equation}
near  which the state is stable. 	
\end{enumerate}
\par

Note that Eq.~(\ref{H}) must not contradict  the rules 1, 2. 
While $x$ changes according to the rules 1 and 2,  the point defined by
Eq.~(\ref{M}), (\ref{H}) form a trace on the $HM$-plane. 
Due to the above postulates the system in our model {\em exactly obeys the return-point memory};
this will be discussed in the next section.
 
\medskip
{\footnotesize
The model admits following interpretation in terms of the ``ensemble of domains''.
We may consider components $x_\alpha$ as indicating the magnetic state of the ``domains'', 
and assume that each of them reverses its sign in one  Barkhausen jump.
In this case for the specimen of unit volume instead of Eq.~(\ref{M}) holds
\[
M=\sum^N_{\alpha=1}m_{\alpha} x_{\alpha}\,,
\]
where $m_{\alpha} > 0$ and $x_{\alpha}$ denote the absolute value of the magnetic moment  and
its direction respectively for each domain. 
}
\medskip

Due to the hysteresis curve symmetry,
it is natural to assume that $M(-x) = -M(x)$ and $H(-x) = -H(x)$, which means that if the state  $x$ is stable in the external field $H$, the 
state $-x$ is stable in the field $-H$.

\begin{figure}
\vskip -1cm
\hskip -0.5cm\vbox{\includegraphics[scale=.55]{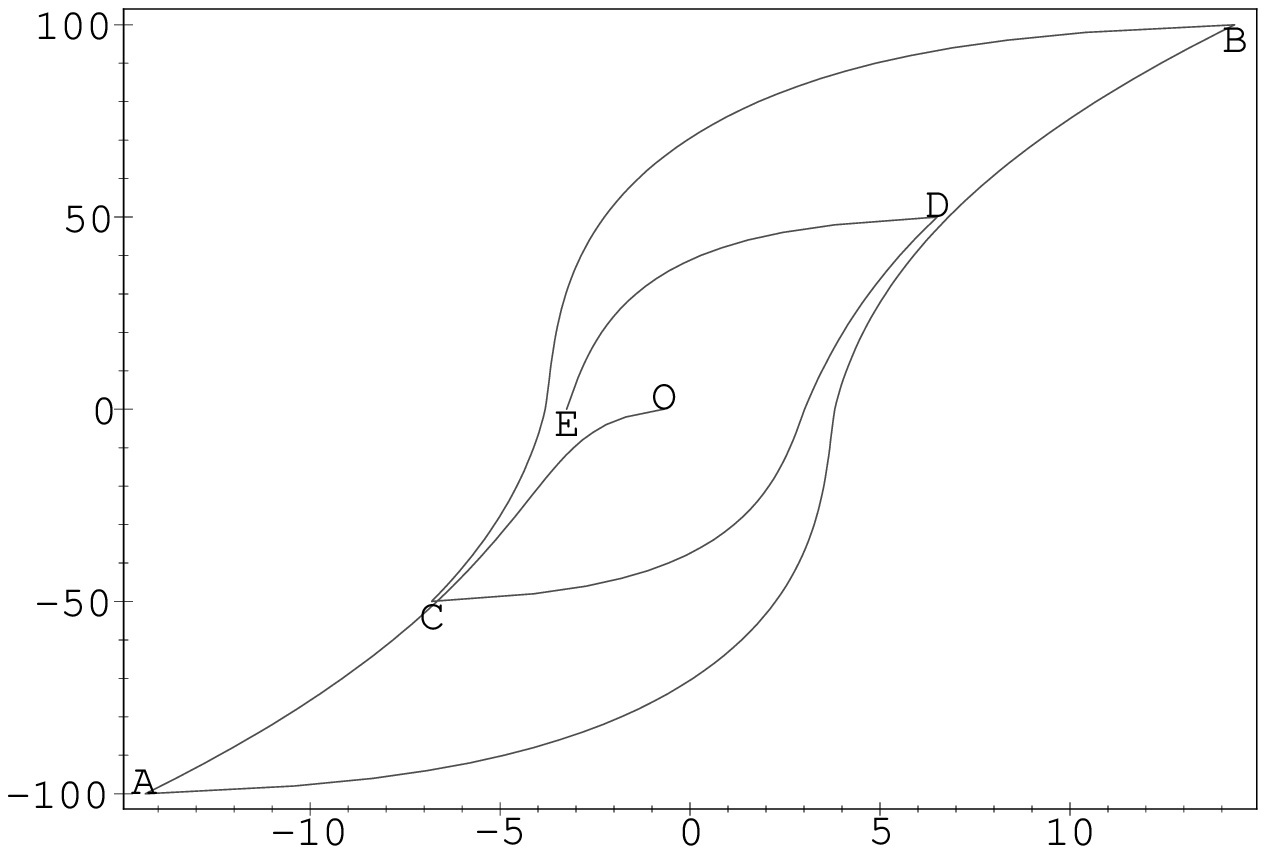} \vskip -1.5cm\includegraphics[scale=.55]{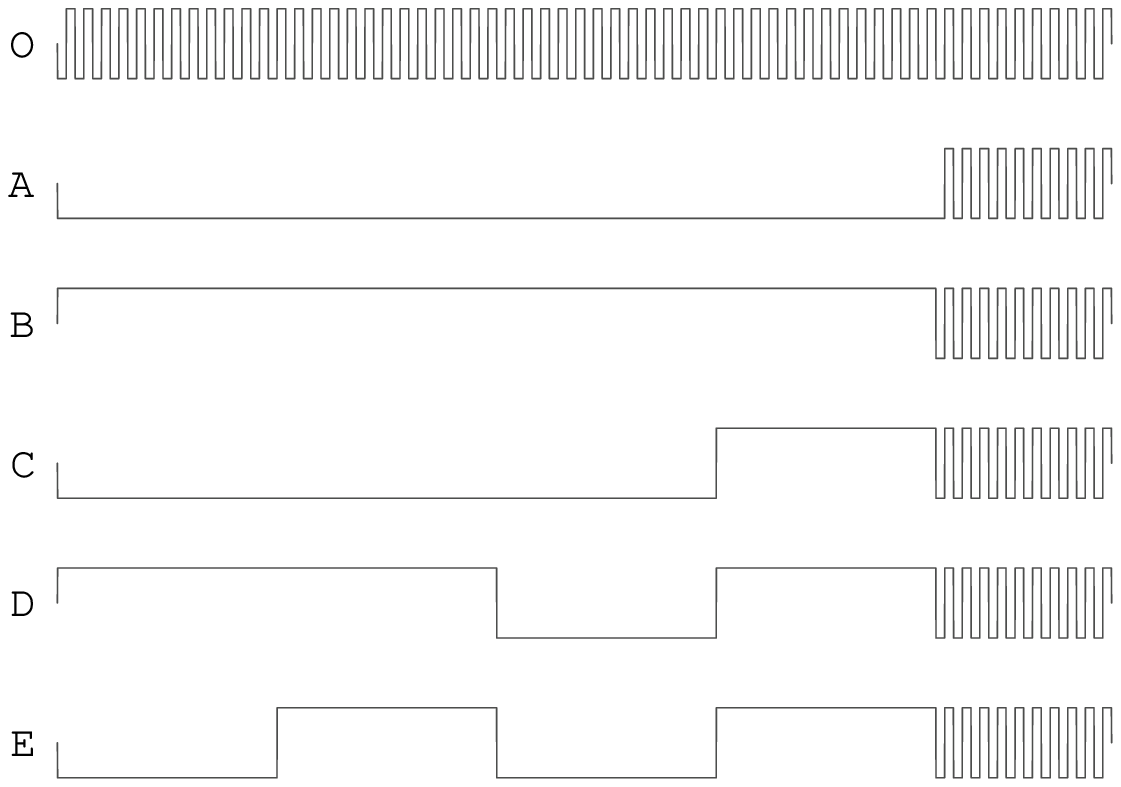}\vskip -1cm}
\caption{
{\footnotesize
State vectors at different points during the magnetization process.
The process starts from the demagnetized state ($O$). On the descending curve  $OA$ the
magnetic field decreases, and $x_\alpha$ become equal to $-1$ one by one from left to right.
After  the point $A$ the magnetic field increases; starting from this point the components $x_\alpha$ become equal to $+1$ one by one,
also from left to right.  After  the point $B$ the magnetic field decreases again and $x_\alpha$
become equal to $-1$ in the same succession,  and so on.
}
}\label{Example}
\end{figure}

All components $x_\alpha$ become equal to $+1$ in a large positive
field, and equal to $-1$ in a large negative field.
Demagnetized state can be obtained as the result of applying to the system alternating
magnetic field of slowly decreasing magnitude, which gives the state vector with ``disordered''
components.
We may take as ``pure'' demagnetized state  the pair of states $(+1,-1,+1, \ldots, \pm1)$ and $(-1,+1,-1,
\ldots, \pm1)$.
The behavior of the state vector is illustrated in Fig.~\ref{Example}.

\bigskip

\begin{figure}
\hskip 0.7cm\includegraphics[scale=0.4]{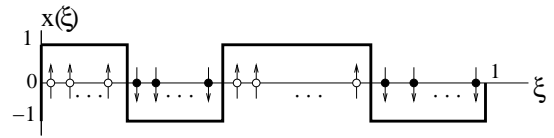}
\caption{\footnotesize 
Illustration of the ``continuous''  variant of the state vector  hysteresis  model. 
Arrows directed up and down correspond to $+1$ and $-1$ components of $x_\alpha$.
}
\label{SVHM-fig}
\end{figure}

In sequel the following ``continuous'' formulation of the model will be convenient.
Let us arrange all  $x_\alpha$ on the interval  $[0,1]$ in the order of increasing $\alpha$  (Fig.~\ref{SVHM-fig}).   
Function $x(\xi): [0,1] \rightarrow \{-1,1\}$  determines the state of the system as follows:
if $x(\xi)$ equals to $+1\;(-1)$ on some interval, then all $x_\alpha$  are equal to $+1\;(-1)$ on this interval.      
Instead of the rules 1 and 2, which determine the behavior of components $x_1\ldots , x_\alpha$, we have the following rules
for $x(\xi)$.

\begin{itemize}
\item[1*. ]  When the magnetic field $H$ increases,  the leftmost point where $x(\xi)$ changes the sign from $+1$ to $-1$,
moves from left to right. 
\item[2*. ]  When the magnetic field $H$ decreases, the leftmost point where $x(\xi)$ changes the sign from $-1$ to $+1$,
moves from left to right. 
\end{itemize}
\par

To express the rules in a simple form, we have assumed that $x(\xi)$ {\em always} changes the sign 
at  $\xi = 0$. 
With $x(\xi)$ the right sides of Eq.~(\ref{M}), (\ref{H}) can be rewritten in the form of functionals
\begin{equation}\label{MHfunc}
 M=M[x(\xi)], \qquad H=H[x(\xi)]. \label{MH}
\end{equation}
Here $H[x(\xi)]$ must not contradict 1*, 2*.

\section{Return-point memory}

Let us show that the model can be obtained as a consequence of the return-point memory \cite{Sethna&all1993}. 

\smallskip
{\em Return-point memory. --- } 
Suppose the system  is evolved under field $H(t)$, where $0<t<T$ and $H(0) \leq H(t) \leq H(T)$ or $H(0) \geq H(t) \geq H(T)$. 
Then for a given initial state of the system  the final state depends only on $H(T)$, and is independent of the time $T$ or the history $H(t)$.
\smallskip

Besides of the return-point memory the existence of states with some special properties is required.
We suppose that a system has two states
$A$ and $B$, with corresponding fields
$H_A$ and $H_B$,  such that: i) the system evolves from the state $A$ to the state $B$, if the magnetic field monotonically increases from $H_A$ to $H_B$;
ii) the system evolves from the state $B$ to the state $A$, if the magnetic field monotonically decreases from $H_B$ to $H_A$.

In the remaining part of this section we will usually assume that the initial state of the system is $A$, 
all inputs $H(t)$ are continuous and piecewise monotonic and for each of them  holds $H(0) = H_A$, $H_A \leq H(t) \leq H_B$.  
We shall restrict our consideration to the set of  states  $\Sigma_{AB}$ reachable from $A$ by applying $H(t)$
that satisfy the above conditions.

Note that if the initial state is $A$, $H(0) = H_A$ and $H_A \leq H(t) \leq H_B$,   
then $H(T)=H_A$ implies that the final state of the system is $A$, and $H(T)=H_B$ implies that the final state of the system is $B$.

\smallskip
{\footnotesize
The last statement concerning the final state $B$  follows directly from the return-point memory. To ensure that it is true for the 
final state $A$ consider the input $\tilde H(t)$ that is applied to the system in the state $B$ and is composed of $H(t)$ before which 
the magnetic field decreases monotonically from $H_B$ to $H_A$. It is clear that $H(t)$ and $\tilde H(t)$ have the same final states; from
the return-point memory follows that the final state of $\tilde H(t)$ is $A$.
}
\smallskip

Consider some input $H(t)$, $H_A \leq H(t) \leq H_B$, $H(0) = H_A$, that is applied to the system in the state $A$.  
Let $H(t)$ increase from $H_0 = H_A$ to $H_1$, then decrease from $H_1$ to $H_2$,
increase from $H_2$ to $H_3$ and so on, where  $H_1, \ldots, H_{n-1}$ are successive maxima and minima of $H(t)$, and $H_n = H(T)$.   

\smallskip

{\em Prime inputs.  --- } Let us call a piecewise monotonic input $H(t)$ prime,
if $H(0)=H_A$, $H_A\leq H(t)\leq H_B$, and one of the following is true: 
(i) the input is invariable $H(t) \equiv  H_A$; 
(ii) the input $H(t)$ is monotonically increasing;
(iii) for all successive maxima and minima holds   
$|H_{k+1}-H_k|<|H_k -H_{k-1}|$ where $k = 1,\ldots , n-1$ and $n \geq 1$.   

It is reasonable do not distinguish prime inputs with identical values $H_0, \ldots, H_n$, because 
according to the return-point memory such inputs give the same final states. 
With this agreement  any prime input is determined by the values $H_0, \ldots, H_n$
where $H_0 = H_A$; the case $n = 0$ corresponds to the invariable input. 

\smallskip
{\em Lemma.  --- } Each state in the set $\Sigma_{AB}$  can be 
obtained from the state $A$ by applying a prime input. 

{\em Proof. --- } 
For any arbitrary state $s$ in $\Sigma_{AB}$  exists piecewise monotonic input $H(t)$ with the final state $s$,
such that $H(0)=H_A$, $H_A\leq H(t)\leq H_B$, $0 \leq t \leq T$. 
If $H(t) = H_A$ at some point $t'$, then $H(t)$ may be replaced on the interval $[0, t']$ with invariable 
function $H(t) \equiv H_A$. 
In the case $t' = T$ we have the invariable input that is prime and has the the final state $A$. 
Otherwise, we may assume that $H(t)$ does not remain constant on any subinterval,
because, as follows from the return-point memory, on each interval of monotony $H(t)$ can be replaced
with monotonically increasing or decreasing function without affecting the final state. 
This means that an arbitrary state $s \neq A$ can be obtained by applying $H(t)$ that increases from $H_0 = H_A$ to $H_1$, 
then decreases from $H_1$ to $H_2$, increases from $H_2$ to $H_3$ and so on, where $H_A <  H_k \leq H_B$ for all $k = 1 \ldots n$, and
$n$ is the number of monotony intervals. 
If $H(t)$ is non-prime, then there are four successive extrema $H_{k-1}$, $H_k$, $H_{k+1}$, $H_{k+2}$ such that
$|H_{k-1} - H_k| > |H_{k} - H_{k+1}|$, $|H_{k} - H_{k+1}| \leq |H_{k+1}$, $H_{k+2}|$. 
According to the return-point memory, $H(t)$ on the interval where $H$ changes from $H_{k-1}$ to $H_{k+2}$ 
can be replaced with monotonic function without affecting
the final state. Such replacement can be repeated not more than finite number of times, because the number of monotony intervals
is finite, and finally we get a prime input with the final state $s$. The lemma is proved.
\smallskip
	
After applying a prime input to the system in the state $A$ some definite state in $\Sigma_{AB}$ is obtained,
which can be assigned to the prime input. 
According to the lemma, such correspondence between prime inputs and states defines the mapping of the set of prime inputs {\em on} the set $\Sigma_{AB}$.

\begin{figure}
\hskip 0.5cm\vbox{\includegraphics[scale=0.5]{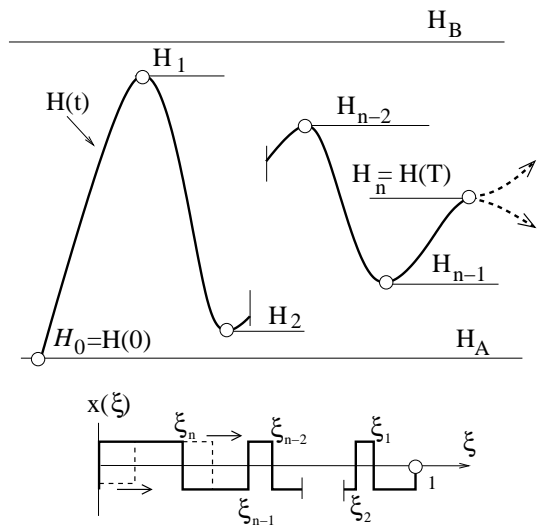}}
\caption{
{\footnotesize
Prime input $H(t)$ and corresponding $x(\xi)$. If $H$ increases or decreases after the final point
(dashed lines) $x(\xi)$ changes according to the rules 1*, 2*. 
}
}\label{Prime_Inputs}
\end{figure}

Let us consider now the set $X$ of functions $x(\xi): [0,1]\rightarrow \{-1,1\}$  which change the sign in a finite number of points.
We can establish one-to-one correspondence between prime inputs and functions $x(\xi)$ in the following way.
Let us assign to the invariable input $H(t) \equiv H_A$ the function $x(\xi) \equiv -1$.
For prime inputs with $n \geq 1$ let us define the points 
\begin{eqnarray}
\xi_1 &=& |H_1 - H_0|/|H_B - H_A|, \nonumber \\
\xi_2 &=& |H_2 - H_1|/|H_B - H_A|,\nonumber\\
&&\cdots \nonumber\\
\xi_n &=& |H_n - H_{n-1}|/|H_B - H_A|.\nonumber
\end{eqnarray}

According to the definition of prime inputs we have $ 0<\xi_n<\ldots \xi_2<\xi_1\leq 1 $.

Let us assign to a given prime input the function $x(\xi)$ that changes its sign
from $+1$ to $-1$ at the point $\xi_1$, from $-1$ to $+1$ at the point $\xi_2$ and so on.
The above determines $x(\xi)$ for any given prime input and vice versa. 
It is illustrated in Fig.~\ref{Prime_Inputs} from which it is clear that $x(\xi)$ changes according to
the rules 1* and 2* of ``continuous'' state vector model, when the magnetic field decreases or increases after
the final point of a prime input.
For the final value of the magnetic field holds

\begin{equation}\label{H0}
H = \frac{H_A+H_B}{2} + \frac{H_B-H_A}{2}\int^1_0 x(\xi) d \xi.
\end{equation}

The right side of this formula is a functional that we denote as $H[x(\xi)]$. 
As the result we have the following proposition.

\smallskip
{\em Proposition.  --- }  The set $X$ of functions $x(\xi)$ can be mapped on the set of states $\Sigma_{AB}$, and $H[x(\xi)]$ exists, 
such that if $x(\xi)$ changes according to the rules 1* and 2*, the state that corresponds to $x(\xi)$ changes    
in the same way as the state of the system under varying external field $H = H[x(\xi)]$.
\smallskip

Once  $x(\xi)$ determines the state of the system, 
its physical properties can be represented  as the functionals with argument $x(\xi)$;
in particular, $M=M[x(\xi)]$. 

As follows from {\em Proposition}, the model is applicable to any system that is compliant to the
return-point memory and has the states $A$ and $B$ with above mentioned properties. 
The last condition is normally satisfied: we can take as $A$ and $B$ the states at the vertices
of the limiting or any minor hysteresis loop.  
Note that $x(\xi)\equiv -1$ corresponds to the state $A$ and $x(\xi)\equiv +1$ corresponds to the state $B$.
For the symmetric hysteresis loop $H_A = -H_B$ and, according to Eq.~(\ref{H0}), $H[x(\xi)]$ is antisymmetric; 
from the symmetry of hysteresis curves with respect to the origin of the $HM$-plane follows antisymmetry of $M[x(\xi)]$.

We have proved the sufficiency of the return-point memory for simulation by the model.
The necessity also can be shown. Namely, it can be proved that any $H[x(\xi)]$ which continuously and monotonically increases and decreases  when
$x(\xi)$ changes according to the rules 1* and 2* correspondingly, is compliant with the return-point memory.
   
\section{Functionals $H[\lowercase{x(\xi)}]$, $M[\lowercase{x(\xi)}]$}

Different forms of functionals $H[x(\xi)]$, $M[x(\xi)]$ are suitable for hysteresis simulation. 
Let us consider some of the simplest. 

\subsection{Linear functionals}
  
Let us try  as $H[x(\xi)]$, $M[x(\xi)]$ linear functionals:
\begin{eqnarray}
M &=& \int_0^1 m(\xi) x(\xi)\,d\xi, \label{intM} \\
H &=& \int_0^1 h(\xi) x(\xi)\,d\xi, \label{intH}
\end{eqnarray}
where $m(\xi)$, $h(\xi)$ are some continuous positive functions.

Consider $x(\xi)$ such that $x(\xi) = -1$ for all $\xi$ on the interval $0 < \xi \leq \xi_1$ for some given $\xi_1$. 
According to the model postulates it means that the magnetic field was decreasing before. 
Let the magnetic field begin to increase, and denote $\xi'$  the point where $x(\xi)$ changes the sign from $+1$ to $-1$, assuming $\xi' \leq \xi_1$.
In this case
\begin{equation}
2\int_{0}^{\xi'}\,d\xi = | \Delta M |, \quad 2\int_{0}^{\xi'} h(\xi)\,d\xi = | \Delta H |, \label{implicit_phi}
\end{equation}
where $\Delta M$ and $\Delta H$ are changes of $M$ and $H$ starting from the origin of new  ascending hysteresis branch;
the same equations are true in the case of any descending branch.

We can see that from Eq.~(\ref{intM}), (\ref{intH})  follow that
all the ascending hysteresis branches and all the descending  hysteresis branches are congruent, and can be 
described with  the same  function $\varphi$, which is determined by Eq.~(\ref{implicit_phi}):
\begin{equation} \label{branch_0}
\Delta H = \pm \varphi \, (\pm \Delta M).
\end{equation}
Here ``$+$'' corresponds to ascending and ``$-$'' to descending branch.

Let us divide the  interval $[0,1]$ into $2n$ equal subintervals, and
define the state $x^{(0)}$ so that  $x^{(0)}(\xi)=-1$ on the odd subintervals and $x^{(0)}(\xi)=+1$ on the even ones.
In accordance with the model postulates we may consider $x^{(0)}$ (or $-x^{(0)}$) as the demagnetized state, 
which is obtained via a demagnetization process performed as the consequence of $n$ demagnetization cycles.
From Eq.~(\ref{intM}), (\ref{intH}) and the continuity of $m(\xi)$ and $h(\xi)$ follow that $H(x^{(0)}) \rightarrow 0$,  
$M(x^{(0)}) \rightarrow 0$ when $n \rightarrow \infty$.
On the initial magnetization curve instead of Eq.~(\ref{implicit_phi}), (\ref{branch_0}) holds
\[
 \int_{0}^{\xi'}m(\xi)\,d\xi = | M | ,   \int_{0}^{\xi'} h(\xi)\,d\xi = | H |,
\]
and
\begin{equation} \label{initial_curve_0}
H = \pm \frac{1}{2} \varphi \, (\pm 2 M).
\end{equation}

Two first terms of Taylor series for $\varphi^{-1}$ in Eq.~(\ref{branch_0}), (\ref{initial_curve_0}) 
correspond to Rayleigh relations  \cite{Bozorth}; this lead us to a conclusion 
that the linear approximation of the functionals is suitable in the case of small fields.

\subsection{One nonlinear expression for $H[x(\xi)]$}

Let us consider as an example following expression
\begin{equation}
H=  a(M)\int_0^1 h(\xi) x(\xi) d\xi + b(M), \label{nonlinearH}
\end{equation}
assuming that $M[x(\xi)]$ is represented by the right side of Eq.~(\ref{intM}).
Here $a(M)$ and $b(M)$ are some functions;
due to the hysteresis loop symmetry $a(M)$ must be even and $b(M)$ must be odd.


\begin{figure}
\hskip -0.5cm\vbox{\vskip-1cm\includegraphics[scale=0.5]{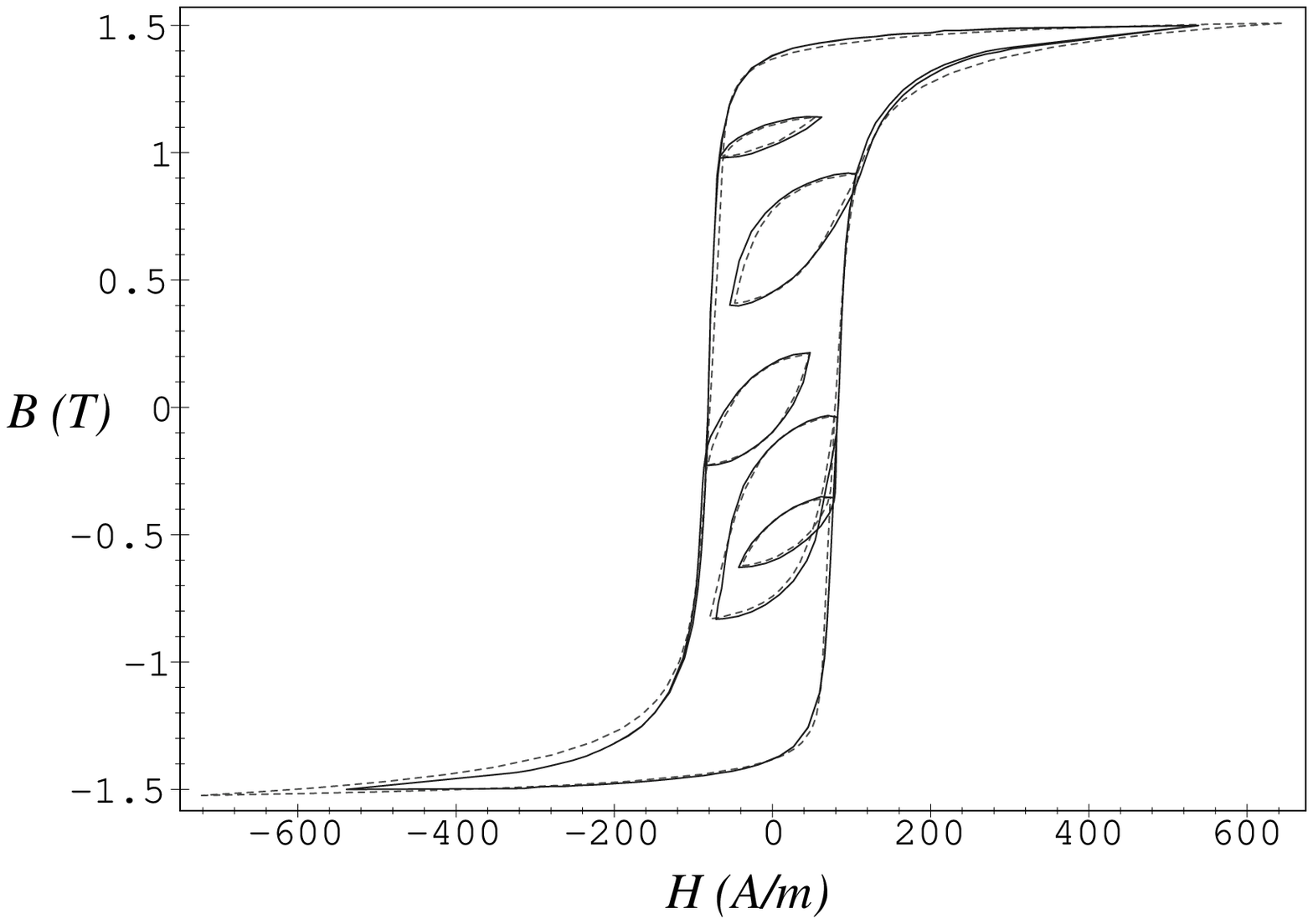}
\vskip-1.5cm\includegraphics[scale=0.5]{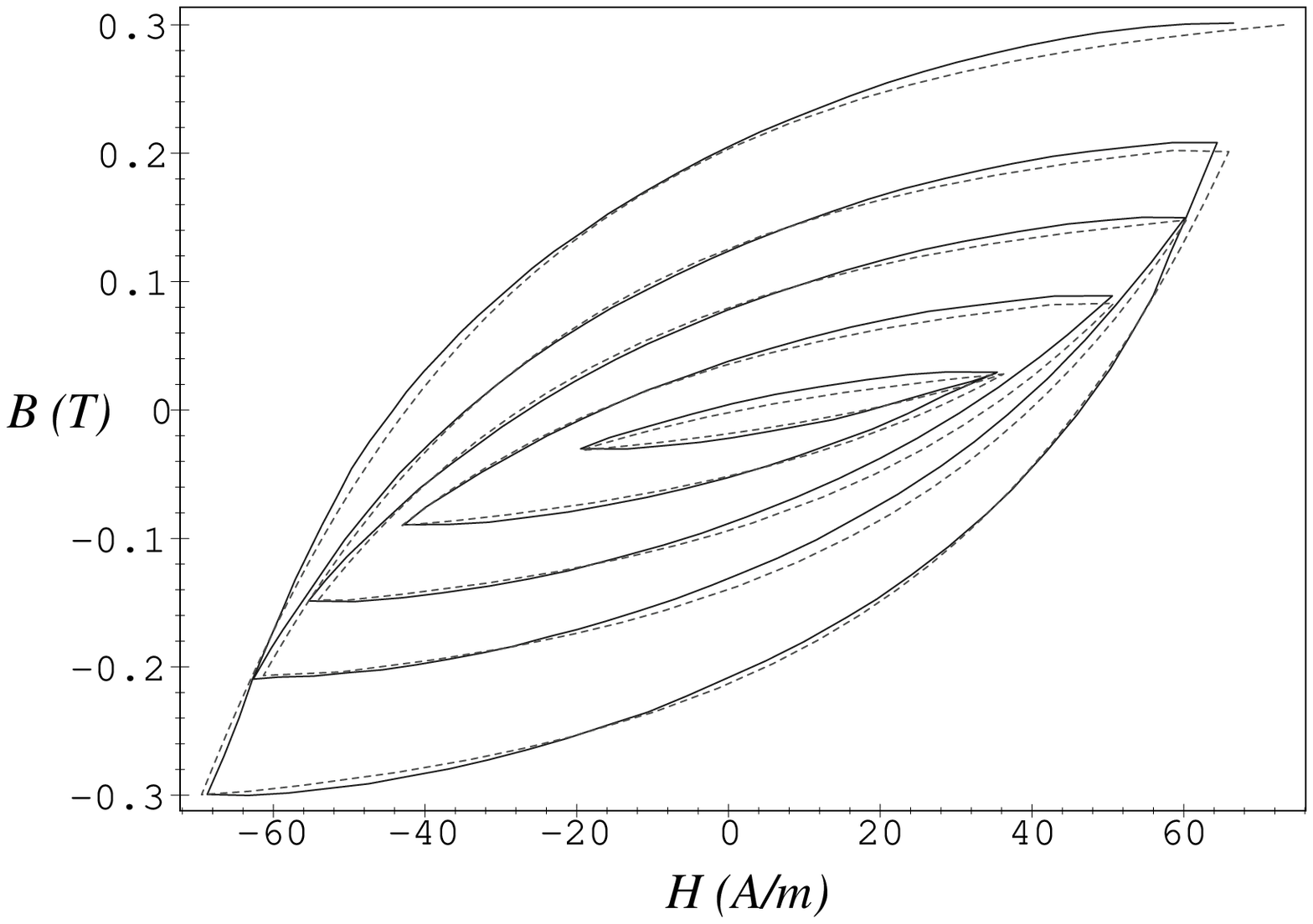}
\vskip -0.5cm
}
\caption{
{\footnotesize
Experimental hysteresis curves of low-alloyed electrical steel (solid lines) and simulated curves (dashed lines).
The simulation is equally accurate in the intermediate and in the small fields.   
}
}\label{Experimental_Curves}
\end{figure}

\medskip
{\footnotesize
From Eq.~(\ref{intM}), (\ref{nonlinearH}) we can found explicit equations for branches of the
hysteresis curve and the initial magnetization curve
\begin{eqnarray}
H &=& \pm a(M) \varphi ( \pm (M - M_0)) +b(M) +a(M)\frac{H_0-b(M_0)}{a(M_0)}, \nonumber \\
H &=&  \frac{1}{2}a(M) \varphi(2M) + b(M) \quad (M>0). \nonumber
\end{eqnarray}
Here $\varphi$ is determined by $m(\xi)$, $h(\xi)$ in the same way as it was considered previously, 
and $M_0, H_0$ denote coordinates of the beginning of a hysteresis branch.
\par}
\medskip

These formulae were applied for hysteresis simulation of low-alloyed electrical steel.
Experimental and simulated curves are shown in Fig. \ref{Experimental_Curves}.

\section{Connection to Preisach model}

The Preisach distribution $p(h_u,h_c)$ describes density of the ``domains'' on the Preisach plane. 
Each domain has a shifted square hysteresis loop;
$h_c$ is the coercive force of the domain and  $h_u$ denotes the shift.    
Function $p(h_u,h_c)$ must be positive, characterized by the symmetry $p(h_u,h_c)=-p(-h_u,h_c)$, and
normalized to unity.

\begin{figure}
\vskip 0.3cm
\hskip 1cm\includegraphics[scale=0.4]{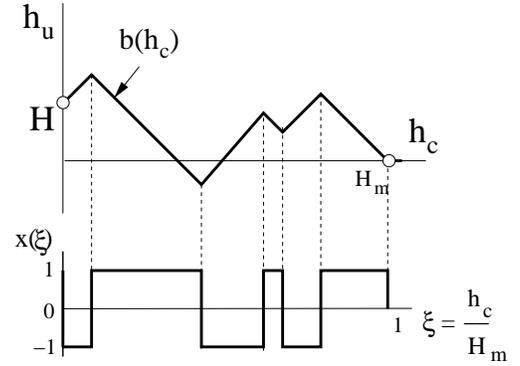}
\vskip 0.3cm
\caption{\footnotesize 
A geometric illustration of the connection between Preisach model and the state vector model. 
}
\label{Preisach}
\end{figure}


Let the magnetic field $H$ always remains in the interval $[-H_m,H_m]$, 
where $H_m$ is as large as desired but fixed; 
we assume also that $p(h_u,h_c)$ is a finite function with support in the triangle 
with vertices $(0,H_m), (0,-H_m), (H_m,0)$. 

Magnetization $M$ is expressed as the integral over all domains
\[
M = M_s\!\!\int\!\!\!\int\limits_{D_{+}}\!\! dh_u dh_c p(h_u,h_c) -  M_s\!\!\int\!\!\!\int\limits_{D_{-}}\!\! dh_u dh_c p(h_u,h_c),
\]
where $D_+$ and $D_-$ are regions with positive and negative domain orientation respectively. 
The boundary $h_u=b(h_c)$ between these regions is a broken line, made of segments with positive 
and negative slope $db(h_c)/dh_c= \pm 1$. The magnetization and the magnetic field are expressed 
via $b(h_c)$:
\begin{equation}\label{HM(b)}
H = b(0),  \qquad M = 2M_s\!\!\!\!\int\limits_0^{\;\; H_m} \!\!\! dh_c \!\!\!\int\limits_0^{b(h_c)}\!\!\! p(h_u,h_c)\, dh_u.
\end{equation}

Let us define
\begin{equation}\label{x}
x(\xi) = -\frac{1}{H_m}\frac{d}{d\,\xi} b\,(\xi H_m) \,, \quad 0\leq\xi\leq 1.
\end{equation}
Function $x(\xi)$ equals to $\pm 1$ and  can determine the state of the Preisach ensemble:
\begin{equation}
b(h_c) = H_m\!\!\int\limits_{h_c/H_m}^1 \! x(\xi) d\xi.
\end{equation}
Now Eq.~(\ref{HM(b)}) can be rewritten in the form of functionals Eq.~(\ref{MHfunc}):
\begin{equation}\label{pHM}
H = \tilde H[x(\xi)], \quad  M = \tilde M[x(\xi)],
\end{equation}
where
\vskip -1cm
\begin{equation}\label{H(x)}
\tilde H[x(\xi)] = H_m\int\limits_0^1\!x(\xi)\,d\xi,
\end{equation}
\vskip -0.3cm
\noindent

and
\vskip -0.5cm
\begin{equation}\label{M(x)}
\tilde M[x(\xi)] = 2M_s\int\limits_0^{H_m} dh_c\!\!\!\!\!\!\!\!\!\!\!\!\!
\int\limits_0^{\;\;\;\;\;\;\;H_m\!\!\!\!\!\!\!\int\limits_{h_c/H_m}^{1}\!\!\!\!\!\!x(\xi)d\xi}\!\!\!\!\!\!\!\!\!\!\!\! p(h_u,h_c)\, dh_u.
\end{equation}

The boundary $b(h_c)$ changes 
in a well-known manner, what can be expressed  in terms of $x(\xi)$. {\em This gives exactly the rules 1*, 2*}, and in conjunction with Eq.~(\ref{pHM}) 
we get the state vector hysteresis model with the functionals of special form defined by  Eq.~(\ref{H(x)}),~(\ref{M(x)}).
Connections between two hysteresis models is illustrated in Fig.~\ref{Preisach}

\medskip
{\footnotesize
The traditional Preisach model has some disadvantages such as 
zero value of the turning point susceptibility and the congruency property \cite{Mayergoyz1986}.
Limitations of the original Preisach model take much weaker form in its modifications. 
It is possible to overcome partially the congruency property \cite{Bertotti} by introducing internal mean field $H_{MF}(M)$,
which depends on the magnetization. This extension of Preisach model corresponds to the functionals
\[
H = \tilde H[x(\xi)] - H_{MF}(\tilde M[x(\xi)]), \quad M = \tilde M[x(\xi)].
\]
Another variant, the product Preisach model \cite{Kadar1999}, also can be represented 
in a form of the state vector hysteresis model. In this case we have
\[ 
H = \tilde H[x(\xi)], \quad M = G(\beta \tilde H[x(\xi)] + \tilde M[x(\xi)]),
\]
where $G$ is ``transformation function'', and the constant $\beta$ provides non-zero value of the turning 
point susceptibility.
}
\medskip

\section{Discussion and conclusions}

The return-point memory is the only essential condition on a system,  that is necessary for simulation by the  model. 
This is an obvious advantage comparative to Preisach model, which requires an extra condition, congruency, not exhibited by 
majority of systems \cite{Mayergoyz1986}.       

Preisach model can be represented as a special case of the model, with
a particular form of the functionals $H[x(\xi)]$, $M[x(\xi)]$. Explicit expressions for
$H[x(\xi)]$, $M[x(\xi)]$ are found in the cases of traditional Preisach model and some of its extensions.
From our consideration follows that a rather complex functional Eq.~(\ref{M(x)}) that corresponds to Preisach model 
is not necessary for hysteresis simulation; it can be replaced with much simpler functionals that may not require 
two-dimensional integration (see as an example Eq.~(\ref{intM}), (\ref{nonlinearH}) and Fig.~\ref{Experimental_Curves}). 

It is worth noting that a general approach for representing {\em states of a system}
was established as a consequence of the return-point memory.
This grants the right to use the model for any physical value $y$ that can be expressed 
as a function of state; corresponding functional $y[x(\xi)]$ determines 
its behavior under varying input.
Note that the model does not comprises any other properties except the return-point memory, 
and compliance with thermodynamics must be considered as a restriction on the functionals.

Different forms of the functionals could be proposed for different kinds of materials.
Actually a class of models is defined, depending on a particular form of the functionals. 
Due to extra flexibility of the model simpler calculations and more precise simulation can be expected,
which may assist in systematizing experimental data.    

\bigskip


\end{document}